\documentclass[
]{elsarticle}

\usepackage{subcaption}

\usepackage[pagewise, modulo]{lineno}

\usepackage{color}
\definecolor{revised color}{rgb}{0, 0.1, 0.8} 

\journal{Chemical Physics Letters}

\begin{document}

\begin{frontmatter}

\title{Modeling Self-Healing Behavior of Vitrimers using Molecular Dynamics with Dynamic Cross-Linking Capability}

\author[mymainaddress]{Gurmeet Singh}

\author[mymainaddress]{Veera Sundararaghavan\corref{mycorrespondingauthor}}
\cortext[mycorrespondingauthor]{Corresponding author}
\ead{veeras@umich.edu, Tel: +1 (734) 615-7242}

\address[mymainaddress]{Department of Aerospace Engineering, University of Michigan, Ann Arbor MI 48109, U.S.A.}
\begin{abstract}

Vitrimers are a special class of polymers that undergo dynamic cross-linking under thermal stimuli. Their ability to exchange covalent bonds can be harnessed to mitigate damage in a composite or to achieve recyclable aerospace composites. This work addresses the primary challenge of modeling dynamic cross-linking reactions in vitrimers during thermomechanical loading. Dynamic bond exchange reaction probability change during heating and its effect on dilatometric and mechanical response are simulated for the first time in large scale molecular dynamics simulations. Healing of damage under thermal cycling is computed with mechanical properties predicted before and after self--healing.

\end{abstract}


\end{frontmatter}

\topskip0pt
\vspace*{\fill} {\color{red}
Please find and cite the published version of the article here:

\bigskip
Gurmeet Singh and Veera Sundararaghavan "Modeling self-healing behavior of vitrimers using molecular dynamics with dynamic cross-linking capability", Chemical Physics Letters 760 (2020) 137966.

\bigskip
DOI =  doi.org/10.1016/j.cplett.2020.137966}
\vspace*{\fill}

\section{Introduction}

Thermoset polymers find applications in fields ranging from robotics, aerospace, automobile, electronics, and batteries either as is or as a part of fibrous composites \cite{Mcbride2019}. There are some challenges in their usage that need to be addressed, for instance, the inability to recycle and re-process due to the irreversibility of the cross--linking bonds and damage evolution in structurally loaded components. Vitrimers are a promising alternative material system that has been recently designed to address these issues\cite{Capelot2012,Zhang2016,Denissen2016,Li2018,Yuan2011}. Vitrimers contain dynamic cross--links that enable them to behave like thermosets at low temperatures and behave like thermoplastics at higher temperatures \cite{vitrimer}. This enables self--healable aerospace composites where damage can be reversed through heating, or recyclable matrix materials where the matrix can be reclaimed after use \cite{strucvit2,strucvit3,Mcbride2019}. Numerous covalent bond exchange mechanisms have been discovered in the recent past such as amines\cite{snyder2020mechanically,denissen2015vinylogous}, transesterification reactions\cite{miao2020chemistry}, Diels-Alder reaction \cite{chen2002thermally}, radical formation\cite{Yuan2011}, etc. Vitrimers have been recently incorporated in carbon fiber composites\cite{Taynton2016,Zhang2016} and nanocomposites\cite{Yang2014} which expands their scope of applications.

Molecular dynamics (MD) models of conventional thermosets have now found use in the aircraft industry. In recent years, these models have demonstrated reliable prediction of the glass transition temperature \cite{LI20112920}, gelation point\cite{OKABE201678}, thermal expansion coefficient \cite{msslepoxy3}, thermal conductivity \cite{VARSHNEY20093378,msslepoxy2}, elastic properties \cite{BANDYOPADHYAY20112445}, and even complete yield surface \cite{vu2015multiscale} without any experimental inputs allowing for computational materials design.  The primary challenge for vitrimers is the presence of temperature dependent reversible cross--link reactions that dynamically alter the mechanical response. 
Exchange reactions have been modelled in the past via methods such as embedding Monte Carlo (MC) moves into molecular dynamics (MD) or fully MD (using specialized three body potentials) or fully MC simulations to simulate bond swaps \cite{ciarella2018dynamics,Oyarzun2018,wu2019dynamics,Smallenburg2013,Sciortino2017}. These simulations have typically employed coarse grained models (bead--spring) that provide high computational efficiency while approximating the mechanical response. For more quantitative modelling, all--atom MD methods are attractive\cite{Yang2016,Sun2020}, however these methods become computationally demanding when simulating slow chemical and mechanical dynamics. Yang et al.\cite{yang2015molecular} modelled bond exchange reactions in all--atom MD by implementing a distance--based reaction cutoff, which greatly accelerates the chemical dynamics. Bonds were created based on the proximity of reacting atoms and the topology was accepted based on the energy of the new bond. However, the simulation had to be started with a low distance cutoff for stability considerations (to avoid large  changes in energies due to initial reactions) and the cutoff was subsequently increased. In this paper, we avoid this issue using an algorithm for chemical reactions based on a pre and post--reaction templates with fixed proximity cutoff \cite{gissinger2017modeling, plimpton1993fast}. Our approach employs an explicit temperature dependence of reaction probabilities drawn from experimental insights. The approach allows, for the first time, modeling of mechanical property changes in vitrimers during thermal cycling above and below topology freezing point ($Tv$) while demonstrating healing of damage and subsequent recovery of mechanical properties. 

\begin{figure*}[t]
\centering 
\includegraphics[trim=1.55cm 3.5cm 1.7cm 2.1cm, clip=true ,width=0.98\textwidth]{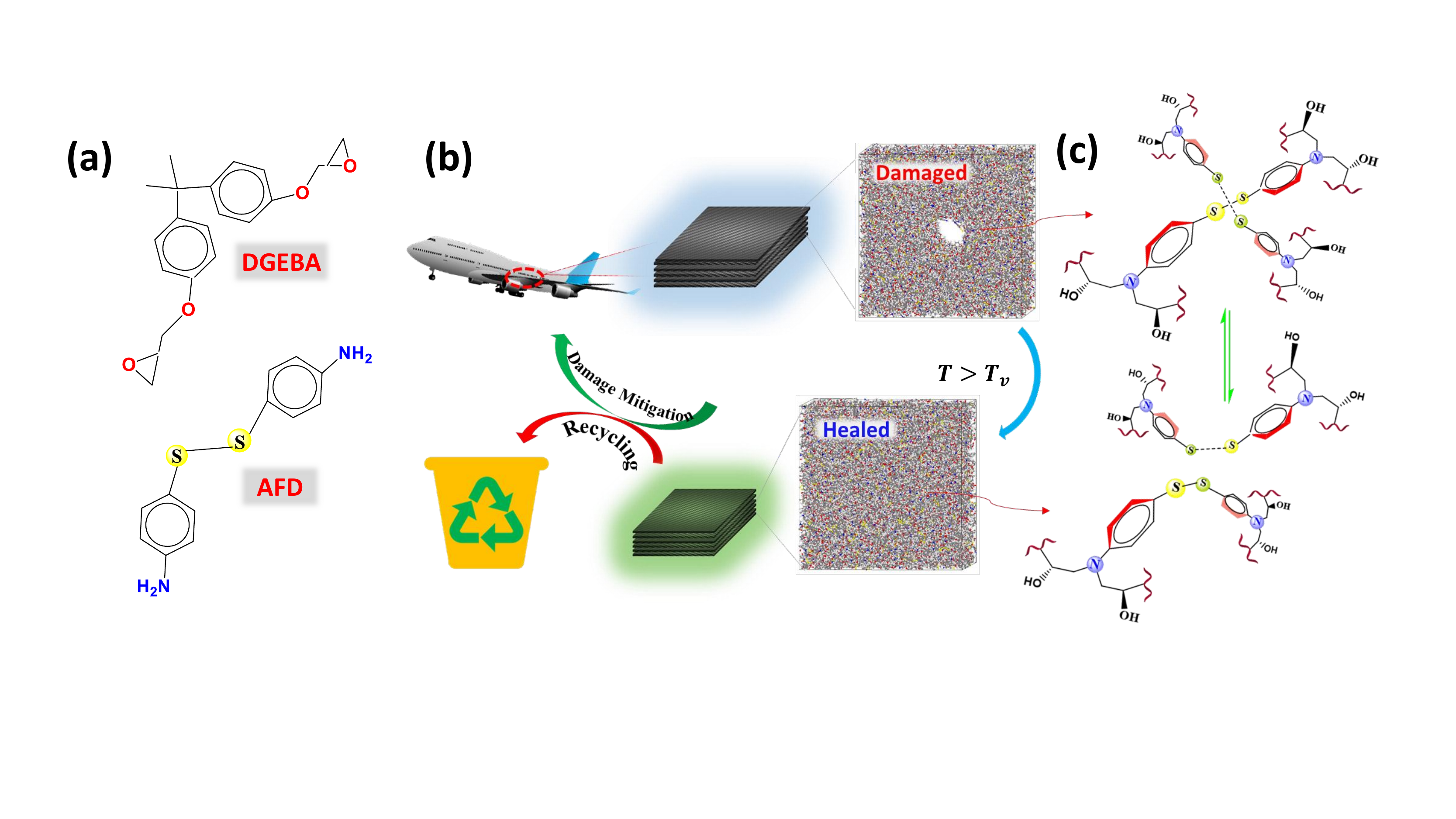}
\caption{(a) Vitrimer monomer units, (b) shows the importance of vitrimer based fibrous composites towards damage mitigation and recycling during operational cycles via (c) the dynamic di-sulfide bond exchange reaction}
\label{fig:intro}
\end{figure*}

In this work, we considered epoxy monomer diglycidyl ether of bisphenol A (DGEBA) cross-linked with 4-aminophenyl disulfide (AFD) vitrimer (structures shown in Fig. \ref{fig:intro}(a)). This particular system was chosen due to its ability to achieve dynamic cross-links in the absence of catalysts\cite{rekondo2014catalyst}. This isomer of AFD also demonstrates mechanochromatic behavior within the visible spectrum (due to the position of amine group at ortho position\cite{de2016transient}) which is useful for damage detection.  Further, this is one of the few vitrimeric systems that has been integrated into a fiber composite \cite{strucvit4}. Through reconfigurable sulfur--sulfur (S--S) linkage (Fig. \ref{fig:intro}(c)), the network can change its topology, preserving the number of bonds but at the same time relaxing its stresses. When the temperature decreases (cooling), the exchange reactions slow down and the network topology appears to be fixed on experimental time scales. Hence, it behaves like an elastic thermoset (elastomer). Thermoset composites are prone to damage during operation and which diminishes their performance (shown in Fig. \ref{fig:intro}(b)). Upon heating vitrimers above the topology freezing point, the dynamic bond exchange reactions accelerate (Fig. \ref{fig:intro}(c)) and the viscosity decreases due to preference towards linear chains, causing the vitrimer to become malleable. Such behavior can be used to heal damage and to recover elastic properties upon cooling back to temperatures below the topology freezing point \cite{de2016transient}. In this paper, we have employed a temperature dependent probability model for the dynamic di-sulfide bond exchange reaction to simulate this effect.

\section{Methods} 

We started with a simulated cell containing a monomer mixture with DGEBA:AFD. The typical synthetic epoxy to hardener stoichiometric ratio of 2:1 was employed \cite{veerapolymer}. We repeat this unit by $8\times8\times8$ to get a simulation box with 2,048 DGEBA and 1,024 AFD  units, with a total of 68,608 atoms. Consistent valence force field (cvff) is assigned to all the atoms with pair, bond, angle, dihedral and improper coefficients modeled\cite{dauber1988structure}. The non-bonded interactions are modeled using Lennard-Jones (LJ) and Coulombic pairwise interaction with a cutoff. The mixture is compressed to a liquid density of $1.0$ gcm$^{-3}$ at 300K and 1 atm NPT. Then the mixture is cured via curing reaction modeling in LAMMPS\cite{plimpton1993fast}. This is not a force-field based reaction modeling, but a bonding procedure for two atoms which mimics a chemical reaction \cite{gissinger2017modeling} and the sites are identified by the pre-- and post-- reaction templates as well as on the mapping between the two templates\cite{gissinger2017modeling} (a schematic shown in Figure S1 of supplemental information (SI)). The primary and secondary amine reactions are modeled with their respective reaction templates as well as their reaction maps (refer Figure S2 of SI). The cut off distance between C and N atoms is set to be 3.5 \AA ~and 5.0 \AA ~for the primary and secondary amine reactions, respectively. In addition to bonding cutoff distance, a reaction probability of one was assigned for the curing reaction. The system is able to achieve up to $95\%$ cross-linking density (Figure S3, S4 of SI). The cured model is then annealed by heating and cooling  cycles at 1K (below $T_g$) and 600K (well above $T_g$) under NPT conditions at 1 bar until the density converges to a $\rho=1.18g/cc$ for neat vitrimer at 1K (refer Figure S5 of SI). The equilibrated structure is used to further study thermo-mechanical properties and self--healing behavior under dynamic S--S bond exchange processes.\\

\begin{figure}[h]
    \centering
    \includegraphics[width=0.75\textwidth]{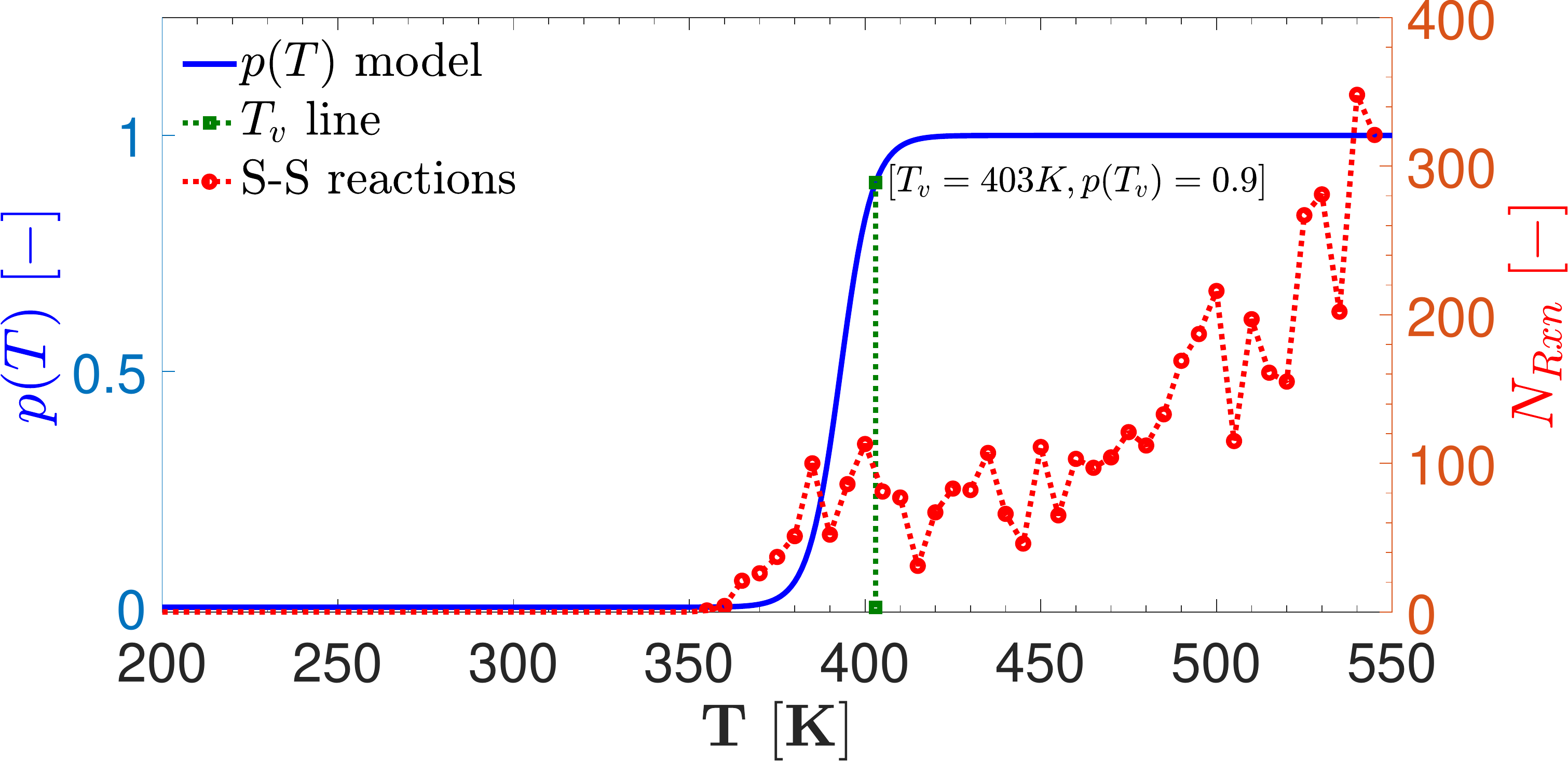} 
    \caption{The reactions probability and the resulting number of S--S bond exchange reactions ($N_{Rxn}$) vs. temperature}
    \label{fig:probmodel}
\end{figure}

The dynamic S--S bond exchange process is modeled in terms of two step reactions. A pre and post-reaction templates are constructed for both  reactions along with a reaction map (refer Figure S6 of SI for reaction templates).  When two pairs of di-sulfide sites come together, they can exchange the chains attached to them as shown in Fig. \ref{fig:intro}(c). The bond exchange reaction can happen when the distance between any sulfur atoms from different chains come within a cutoff distance of $4.12$ \AA ~ (double of the S--S bond length\cite{Steudel1975}, $2.06$ \AA) and when such sites are identified, the probability of the reaction is modeled as a function of temperature. For vitrimers, the temperature at which the acceleration of the reaction rate occurs is referred as topology freezing temperature ($T_v$) and this value can be different or close to the glass transition temperature ($T_g$) of the vitrimer\cite{Denissen2016,van2020vitrimers}. In this paper, we assumed both the transitions occur at the same temperature ($T_v=T_g = 403$K). However, in case of different transition temperatures, the vitrimer behavior can be modeled by considering $T_v$ as the reference for dynamic bond exchange reaction probability. It is observed in experiments that the dynamic bond exchange reactions are very slow at room temperature but can accelerate near or above the topology transition ($T_v$) temperature of this vitrimer system \cite{rekondo2014catalyst, de2016transient}. The modulus is seen to soften and the mobility of the chains accelerate around the transition temperature in a sigmoidal manner\cite{Zhou2019}. Based on this insight, we model the topology transition by accelerating the dynamic bond exchange reactions in a sigmoidal manner around the vitrimer transition temperature. Note that the actual experimental time scales of the exchange reactions are slower\cite{Yu2018}, and accelerated rates serve to realistically capture the thermomechanical behavior within the time scale of molecular dynamics simulations\cite{Yang2016,Sun2020}. To model the onset of dynamic exchange reaction phenomenon in the current model, we assign the probability of S--S bond exchange reaction as a function of the temperature as shown in Fig. \ref{fig:probmodel}.

The probability of the dynamic bond reaction as a function of temperature is given by Equation \ref{prob}.
\begin{equation}\label{prob}
    p(T) = \frac{1}{\left[exp\left( -a\left(T-T_v \right) - ln\left( \frac{p_{T_{v}}}{1-p_{T_{v}}} \right) \right) +1 \right]}
\end{equation}
where, $a=\frac{{2p_{T_{v}}}}{w(1-p_{T_{v}})}$ is determined by  $w$ which is the measure of the width of the transition from glassy to rubbery phase measured, for example, using the modulus vs temperature response (we consider, $w=20$K), and $p_{T_{v}}=0.9$ is the probability of the dynamic bond exchange reaction at $T_v$ ($T_v$ line in Fig. \ref{fig:probmodel}), we assume that at $T_v$, most of the transition starts occurring but it is not complete hence $p_{T_{v}}=0.9$. And, we assume that at the beginning of the transition window, $T_v-w$, the probability is $p({T_{v}-w})=1.0-p_{T_{v}}=0.1$. This is considered in order to obtain a smooth increase in the reaction rate near $T_v$. 

\section{Results and Discussion}

We first analyze the molecular dynamics model without dynamic S--S bond (indicated in plots as `static' model) and with dynamic S--S bond exchange modeled. The annealed structure is equilibrated for $150$ps \cite{Gupta2013} at each temperature starting from $200$K to $550$K at an increment of $5$K. The change in volume is computed and normalized with respect to the initial volume at $200$K ($V_0$). The algorithm keeps a cumulative count of bond exchange reactions that occurred in the system\cite{gissinger2017modeling}. In this system, the number of bonds remain conserved, and hence, bond breaking is accompanied with new S--S bond formation (as sketched in Fig.\ref{fig:intro}(c)). The number of S--S bond--exchange reactions occurring at each temperature is depicted as a  red line in Fig. \ref{fig:probmodel} alongside the chosen reaction probability (in blue). This plot indicates that the number of reactions begin to increase as we get close to $T_v$ and stabilizes around $T_v$. At temperatures well beyond $T_v$ (when the probability of reaction is $1.0$), the number of reactions increase further with increase in temperature. This is attributed to an increase in the frequency of collision events of bonding atoms at higher temperatures. These features are achieved by developing a reaction probability centered around the topology freezing transition point based upon experimental insights\cite{rekondo2014catalyst, de2016transient}. If the vitrimer transition temperature were lower than the glass transition temperature, then the reaction rates will be lower owing to higher stiffness of the polymer which will significantly decrease the collision frequency. As an example, we find that the number of exchange reactions over 150 ps was $53\%$ lower for $T_v$ set at 200K vs 403K for this polymer.  By changing the $T_v$ value, we expect the model to capture the topology transition behavior for other such vitrimers as well.

\begin{figure}[h]
    \centering
    \includegraphics[width=0.75\textwidth]{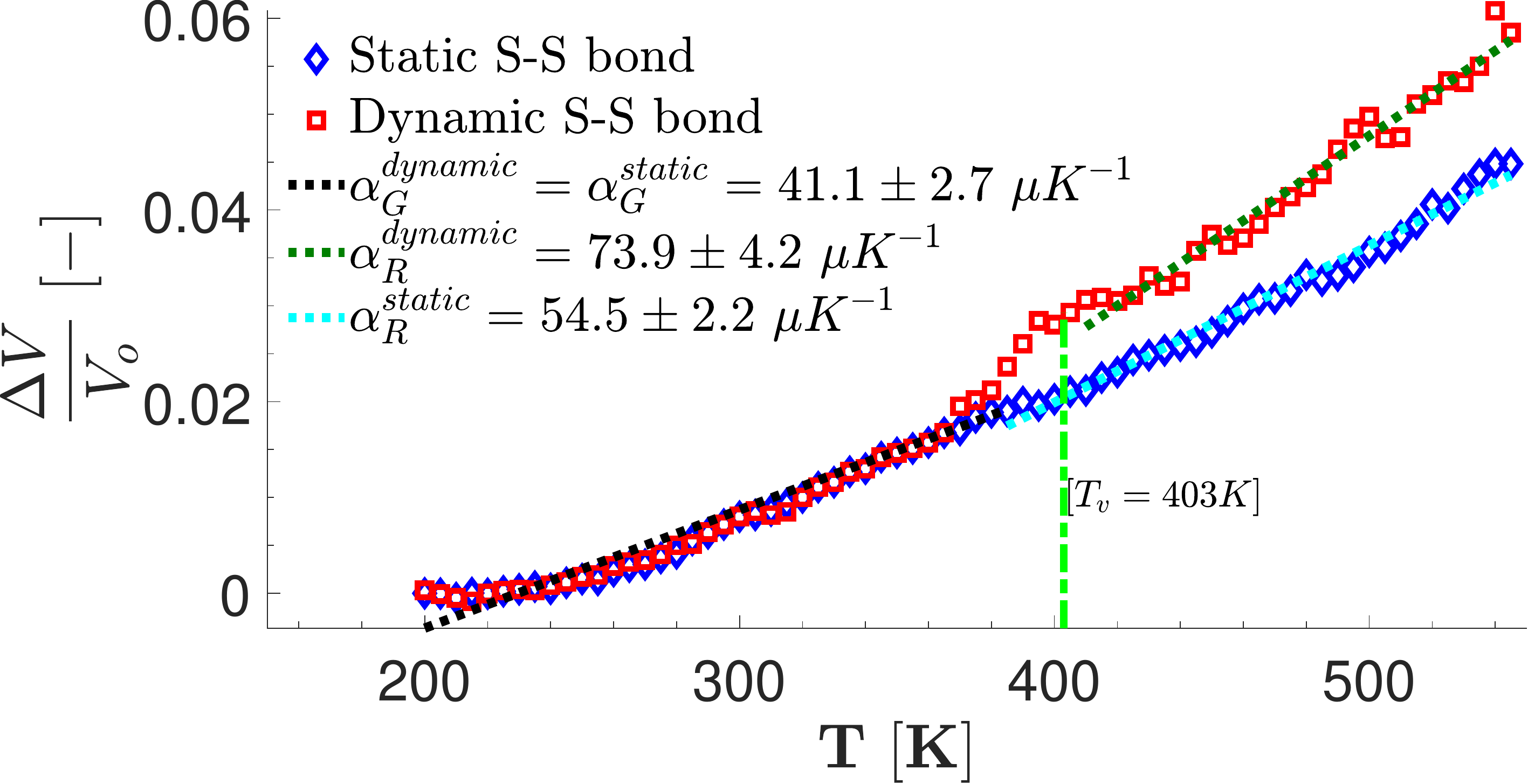} 
    \caption{Volumetric expansion vs. T for model with and without modeling S--S bond reactions. R and G represent rubbery and glassy phases of the glass transition, respectively.}
    \label{fig:Tg}
\end{figure}

Fig. \ref{fig:Tg} shows the V-T characteristic of the neat vitrimer for static and dynamic S--S bond modeled in the simulation. The plot clearly shows that there is no change in the volume expansion for dynamic and static S--S bond until the reactions start picking up near $T_v - w=403$K$-20$K$~=383$K when the probability of the reaction is $1-p_{T_v}=0.1$. We also observe that the dynamic S--S bond model has a higher coefficient of thermal expansion as compared to the static S--S model in the rubbery region beyond $T_v$. This is expected due to the added mechanical flexibility owing to dynamic rearrangements of chains\cite{Capelot2012}. The static coefficient of thermal expansion below $T_v$ of 41.1 $ \pm 2.7 \mu K^{-1}$($=\frac{\Delta V}{3V_o}$) falls at the lower end of range reported ($45.0 - 65\mu ~K^{-1}$) for cured epoxy resins~\cite{boulter1997high}. The model is next employed in a first attempt in literature, to the best of authors knowledge, to understand the effect of dynamic bond exchange reaction on mechanical behavior at different temperatures.

\begin{figure}[h]
    \centering
    \includegraphics[width=0.75\textwidth]{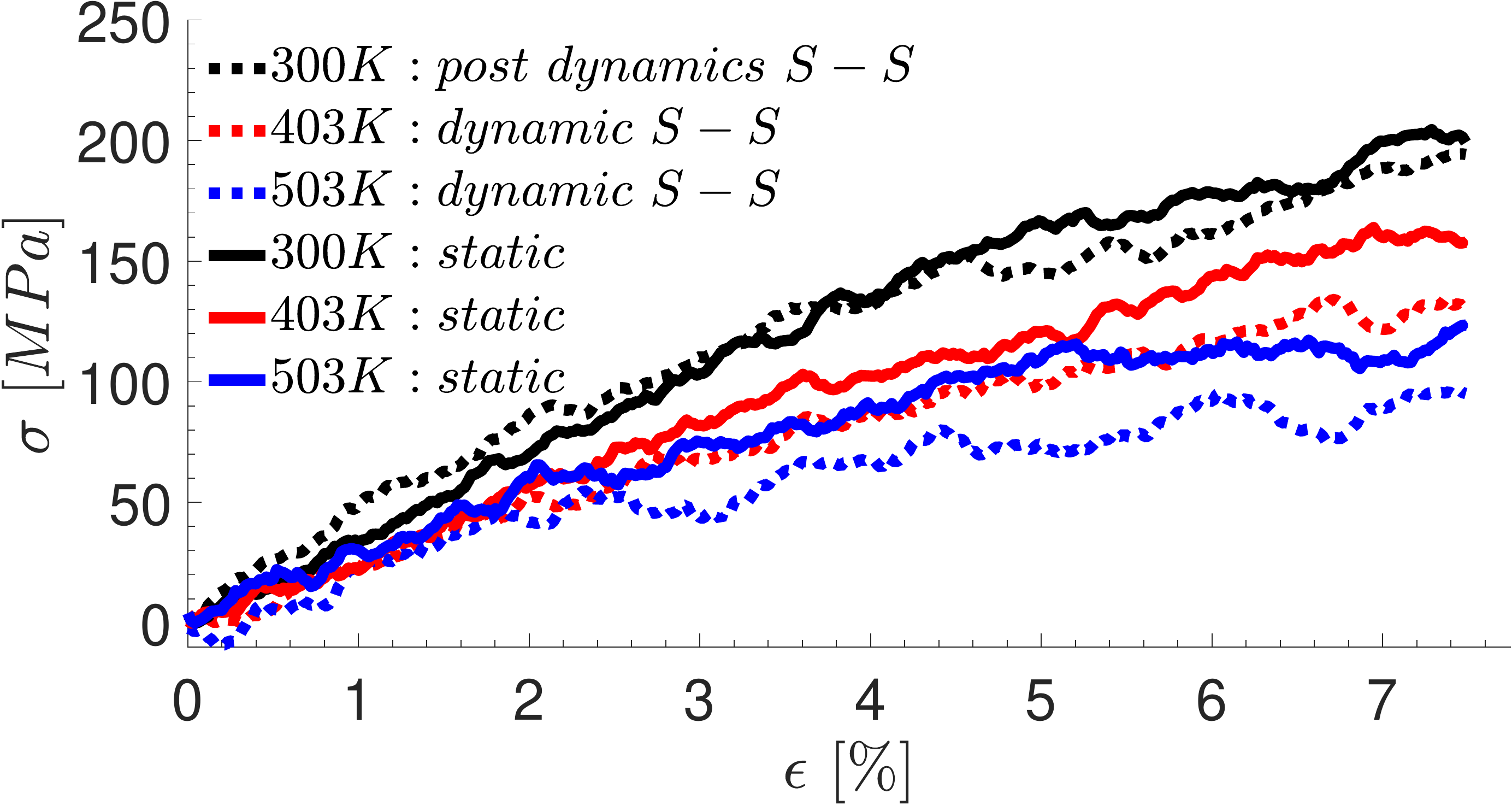}
    \caption{Influence of dynamic S--S bond modeling under uni-axial tension}
    \label{fig:sig-epsilon} 
\end{figure} 

In Fig. \ref{fig:sig-epsilon}, stress--strain relationship  as a result of stretching the simulation box along an axis at a strain rate of  $2\times 10^{8}s^{-1}$ at different temperatures is shown. At higher temperatures, both models show softening behavior. However, the difference in the stress-strain behavior is pronounced at high temperatures ($T_v+100$K) where dynamic bond exchange dominates and vitrimeric material remains softer. The dynamic bond exchange model is taken to $503$K for $250$ps and relaxed back to $300$K to compare the room temperature behavior after thermal cycling. In this case, the post dynamic bond and the static S--S bond cases follow each other well, which shows that despite the model having undergone dynamic bond exchange reactions upon heating, the elastic response is retained at room temperature (300K). This is inline with the understanding of the response of the recycled vitrimers\cite{Denissen2016,Li2018}. Note that due to inherently high strain rates employed in molecular dynamics, the Young's modulus of DGEBA:AFD vitrimer remains overpredicted ($4.5 \pm 0.21$ GPa) compared to experiment ($2.6$ GPa) under quasi-static loading\cite{strucvit4} conditions at 300K. MD simulations have predicted in the range of 3.4 GPa to 5.8 GPa for epoxies\cite{ionita2012multiscale,shokuhfar2013effect,knox2010high, barton1997application} in literature. Note that while a vitrimer behaves like a viscous fluid beyond the topology transition temperature ($T_v$)\cite{Zhang2016,Taynton2016,Yang2014}, molecular dynamics results show sustained stresses beyond $T_v$ due to the high loading rates (as is also seen in MD literature \cite{Yang2016,Sun2020}).

\begin{figure}[h]
    \centering
    \includegraphics[width=0.75\textwidth]{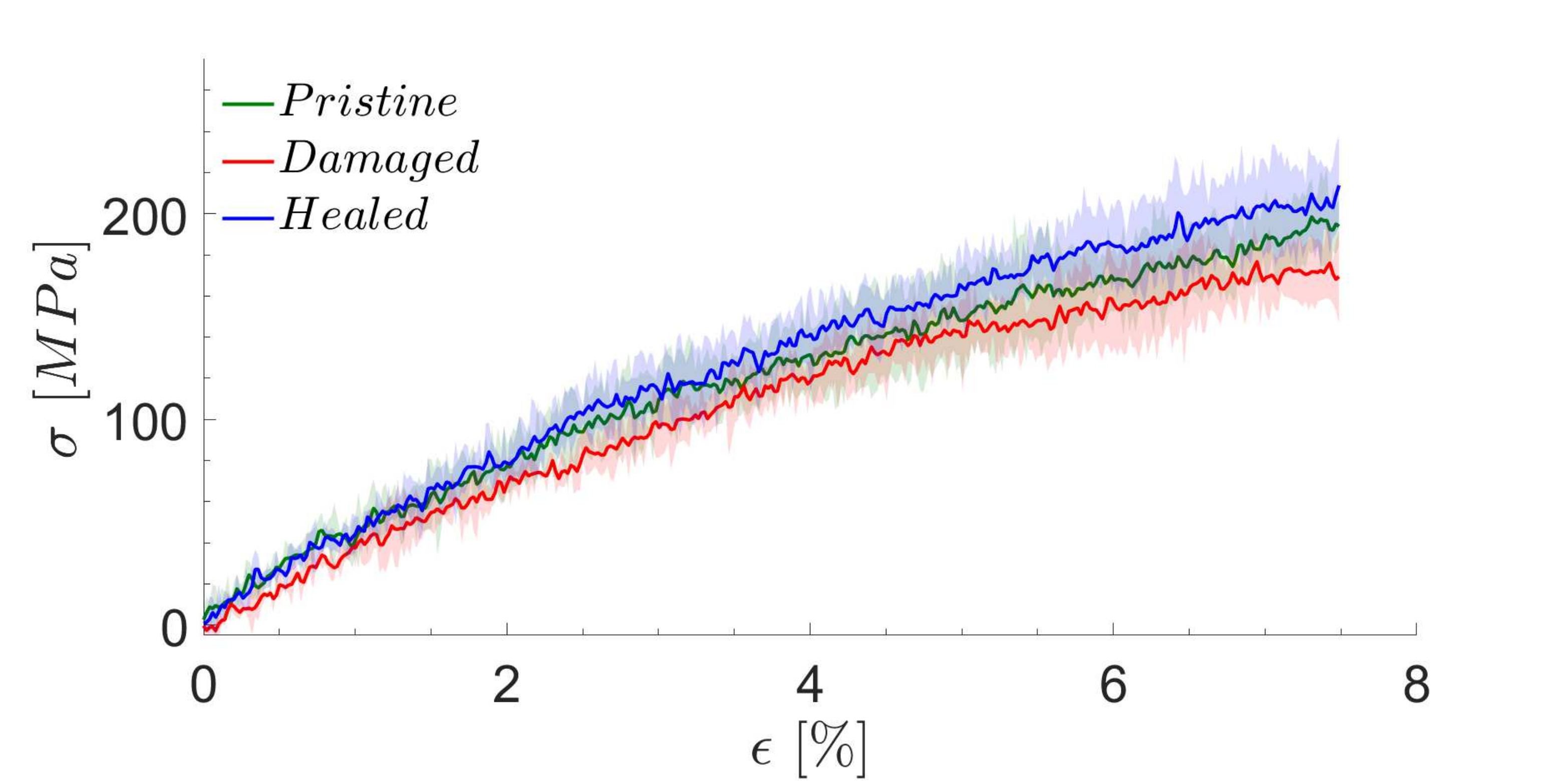}
    \caption{Stress-strain response under uniaxial tension in different directions at 300K for damaged and healed samples.}
    \label{fig:sig-Healing} 
\end{figure}

\begin{figure}[h]
    \centering
    \includegraphics[width=0.75\textwidth]{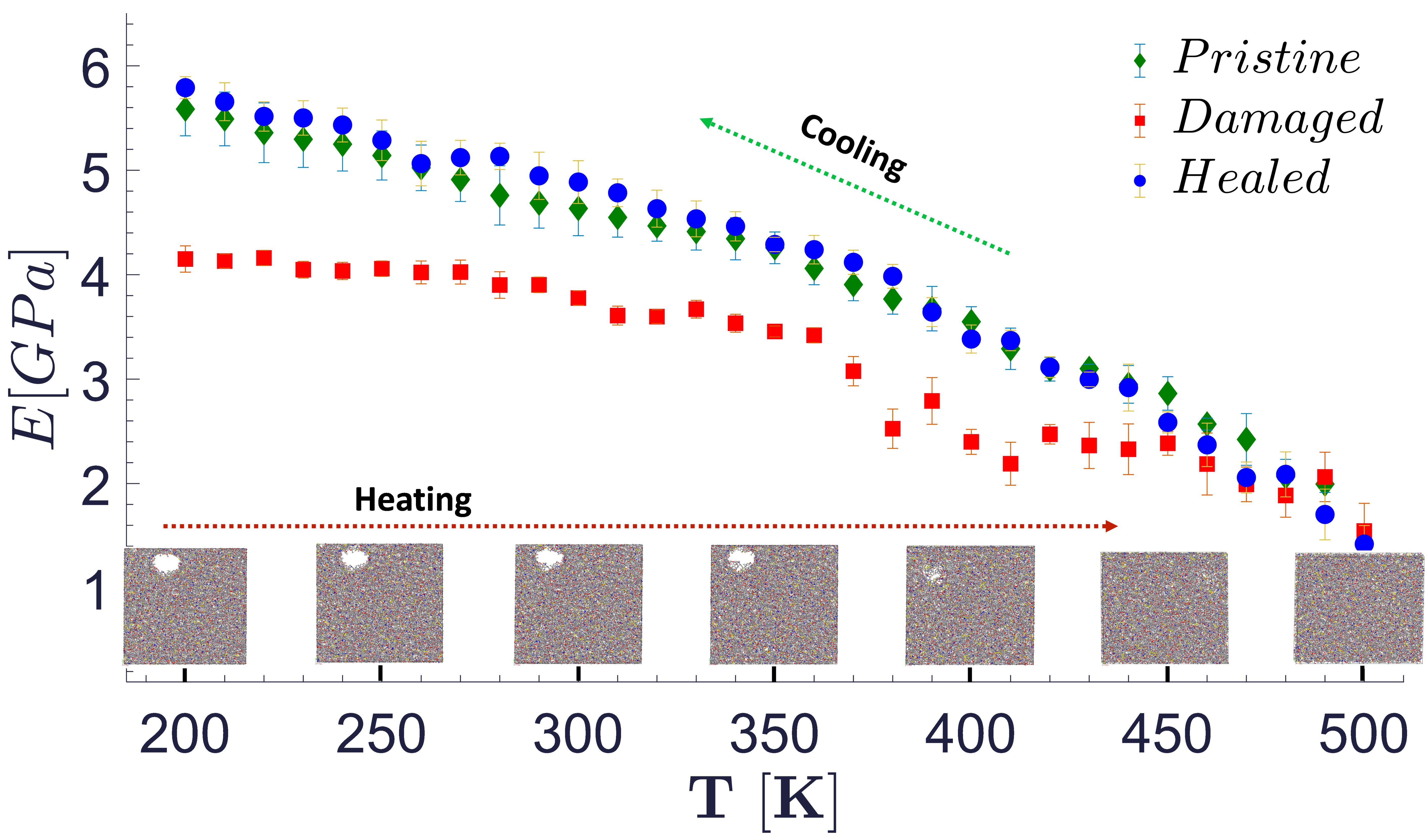}
    \caption{Stiffness enhancement post damage healing via S--S dynamic bond exchange reactions.}
    \label{fig:sig-Healed-Stiffness} 
\end{figure}

We now exploit the dynamic S--S bond exchange reaction capability to demonstrate healing of vitrimers in MD simulations. The damage corresponds to a carbon nanotube (CNT) pullout from the matrix\cite{zhang2017carbon,barber2004interfacial,tsuda2011direct}. We first insert a (12,12) single walled CNT of 91\AA~ length along $z-$axis, and displace the atoms radially in $xy$ plane. The rest of the curing protocol, described in the methods section, remains same as the neat vitrimer model with a final density of $\rho=1.19$~gcm$^{-3}$ at 1K (refer Figure S5 of SI). To generate the damage, the CNT was removed out of the equilibrated simulation box.  Then the simulation box is heated up to a temperature $T_v+100$K$=503$K for $250$ps. The hole heals under the influence of dynamic S--S bond exchange reactions. The healed system is then relaxed back to analyze its elastic response at room temperature.  Fig. \ref{fig:sig-Healing} shows the elastic response of the damaged and healed vitrimers (averaged of three different direction stretches with the bounds shown by light colors) at $300$K. We observe that the stress--strain response of the healed vitrimer is consistently higher than that of the initial damaged sample along all three loading directions demonstrating healing. In order to plot the elastic properties over a range of temperatures, we computed the stiffness by linear regression in $\epsilon=4\%$ range and the elastic modulus was averaged in all three directions. Fig. \ref{fig:sig-Healed-Stiffness} demonstrates that the healed structure is able to recover the pristine vitrimer elastic modulus over the range of the temperatures below topology freezing point. Snapshots of damage healing showing the hole left by CNT pullout being filled via a dynamic bond exchange mechanism post $T_v$ are shown as inset (refer animation in SI). 

\break
\section{Conclusions}

In conclusion, a vitrimer's ability to dynamically reform covalent bonds during thermal cycling allows one to achieve superior behavior over traditional thermosets such as damage healing. To model this behavior, we have developed a novel temperature dependent reaction probability which is integrated with molecular dynamics modeling of vitrimers. The reaction probability is empirically based on the observed reaction kinetics of dynamic bond exchange near or above topology freezing transition temperature ($T_v$). It is shown that the model captures the onset as well as the increase in number of reactions above this transition point owing to higher mobility of chains, without the need to alter reaction distance cutoffs. The vitrimer is seen to achieve softer behavior around and beyond topology freezing transition temperatures while maintaining the glassy behavior at pre-transition temperatures upon thermal cycling. In the simulation of the healing of a cylindrical pore created due to  CNT pullout, the model is able to show both the healing of the vitrimer and complete recovery of elastic modulus upon cooling. Such modeling capability can be further used to achieve insights into the interplay of mechanics and chemistry in a variety of other dynamic bond exchange materials.

\section*{Acknowledgement}
This research was supported in part through computational resources and services provided by Advanced Research Computing at the University of Michigan, Ann Arbor. The authors would like to thank Mr. Siddhartha Srivastava for his valuable discussions.

\section*{Supplementary Information}
Supplementary material contains additional data on curing of the polymer mixture, the annealing protocol, and pre and post--reaction templates of the dynamic bond exchange reaction and an animation of damage healing via dynamic bond exchange.

\noindent \textit{Data Availability Statement:} The data that support the findings of this study are available from the corresponding author upon reasonable request.


\bibliography{mybibfile}

\end{document}